\documentclass[reprint, amsmath,amssymb,aps,prb]{revtex4-1}

\usepackage{graphicx}
\usepackage{dcolumn}
\usepackage{bm}
\usepackage{braket}

\begin{document}


\title{Goldstone modes in the emergent gauge fields of a frustrated magnet}

\author{S. J. Garratt}
 \email{samuel.garratt@physics.ox.ac.uk}
\author{J. T. Chalker}%
\affiliation{Theoretical Physics, University of Oxford, Parks Road, Oxford OX1 3PU, United Kingdom}

\date{\today}

\begin{abstract}
We consider magnon excitations in the spin-glass phase of geometrically frustrated antiferromagnets with weak exchange disorder, focussing on the nearest-neighbour pyrochlore-lattice Heisenberg model at large spin. The low-energy degrees of freedom in this system are represented by three copies of a ${\rm U}(1)$ emergent gauge field, related by global spin-rotation symmetry. We show that the Goldstone modes associated with spin-glass order are excitations of these gauge fields, and that the standard theory of Goldstone modes in Heisenberg spin glasses (due to Halperin and Saslow) must be modified in this setting.
\end{abstract}

\pacs{Valid PACS appear here}
\maketitle

\section{Introduction} 
Gauge fields arise as low-energy degrees of freedom for frustrated magnets in a variety of contexts \cite{read-sachdev, Isakov2004, kitaev,balentsreview}.
Their emergence is particularly transparent in the classical limit, where the systems of interest have macroscopically degenerate ground states and ground-state spin configurations can be mapped to configurations of a divergenceless vector field \cite{Isakov2004}. 
An important application of these ideas has been in research on spin-ice materials, represented by the Ising antiferromagnet on the pyrochlore lattice \cite{Harris1997,bramwell-gingras-review}.
In spin ice, the emergent gauge field is described by a ${\rm U}(1)$ theory, and magnetic monopole excitations act as its sources and sinks \cite{castelnovo-moessner-sondhi}.
Extending the approach to $n$-component classical spins, a distinct flavour of ${\rm U}(1)$ field arises from each spin component: global spin rotations act as rotations between these flavours, and magnetisation density is a vector source for flux. 

Such gauge fields may acquire dynamics by a number of different routes. Starting from a classical model, a natural step is to introduce quantum tunnelling between pairs of ground states that are related via rearrangement of small numbers of spins \cite{moessner-sondhi-tlim,balents-girvin}.
In quantum versions of spin ice this leads to a theory of the standard form familiar from quantum electrodynamics \cite{moessner-sondhi,balents}.
An alternative is to build on the precession of the spins in the exchange fields arising from their neighbours. In Heisenberg models the global rotational symmetry then gives rise to theories of gauge fields with a conserved vector charge. A treatment of the Heisenberg antiferromagnet on the pyrochlore lattice that combines precessional dynamics with a description in terms of emergent gauge fields is appropriate under two conditions: the system should be at low enough temperatures that it is close to its ground-state manifold, but not at such low temperature that quantum order-by-disorder \cite{henley,hizi} establishes the Ne\'el state. Both conditions are satisfied in a window below the Curie-Weiss temperature that is wide at large spin.
In this temperature range the gauge field dynamics arising from precession is overdamped, with a relaxation rate that is predicted \cite{Moessner1998a,Moessner1998,conlon-chalker}
and observed \cite{broholm}
to be proportional to temperature.

Quenched exchange randomness offers a way to explore this physics further. It leaves the gauge fields as distinct degrees of freedom if its amplitude $\Delta$ is much less than the mean exchange $J$, and it enables the system to evade order-by-disorder, instead stabilising the spin-glass state below a freezing temperature $T_{\rm F} \sim \Delta $
\cite{Bellier-Castella2000,Saunders2007,Andreanov2010}. 
The frozen state corresponds to a particular gauge field configuration
(selected by the exchange randomness), which
spontaneously breaks symmetry under global spin rotations. Consequently it offers a platform to  understand the role of the global rotational symmetry in the gauge field dynamics. A key question is whether excitations in this state can be viewed both as excitations of the gauge degrees of freedom and also as Goldstone modes. In this paper we establish a theoretical treatment of these modes, demonstrating how the two perspectives are consistent in the low-energy limit.

The theory of Goldstone modes in conventional spin glasses was established some time ago in work by Halperin and Saslow \cite{HalperinB.I.;Saslow1977}, and by Ginzburg \cite{Ginzburg1978}, and describes the interplay between smooth rotations of the spin configuration and a conserved magnetisation density. Within that approach, the long-distance properties of the ordered state are characterised by the uniform magnetic susceptibility $\chi_0$ and the long-wavelength spin stiffness $\rho$. Modes of frequency $\omega$ and wavevector $k$ have a linear dispersion relation $\omega=ck$ with speed $c=\sqrt{\rho/\chi_0}$. For a spin glass having nearest-neighbour interactions with mean strength zero and variance ${\cal J}^2$, one has $\chi \sim {\cal J}^{-1}$ and $\rho \sim {\cal J}a^{2}$, where $a$ is the lattice spacing. As a result, $c\sim {\cal J} a$. A direct attempt to extend the conventional theory to the spin-glass state in geometrically frustrated Heisenberg antiferromagnets with weak disorder suggests the result $c\sim a \sqrt{J\Delta}$. If correct, this form would imply that the modes mix gauge fields with high-energy degrees of freedom, since it combines ${J}$ and $\Delta$. 

We show below that this is not in fact the behaviour in the pyrochlore antiferromagnet. Instead, the results of our investigation in the regime $0 < \Delta \ll J$ are as follows. We find that fluctuations of the emergent gauge fields acquire a stiffness of order $\Delta$, and that the lowest energy degrees of freedom are smooth rotations of the frozen state. Further, we show that the low-frequency magnons associated with these rotations have speed $c \sim a\Delta$, where $a$ is the lattice spacing. Since this is independent of $J$, it is at odds with the conventional theory. To expose the new features of  the spin glass state in geometrically frustrated  magnets, we develop a continuum description of the low-frequency magnons, thereby extending the hydrodynamic theory to these systems. Our modified hydrodynamic theory reveals that, in the limit of vanishing exchange randomness, the low-energy modes involve only the emergent gauge fields.

A number of geometrically frustrated antiferromagnetic materials are well-described in a first approximation by the Heisenberg model. Many of them show spin freezing with a transition temperature much smaller than the dominant interaction scale (which is characterised by the Curie-Weiss constant) and this freezing is plausibly attributed to weak exchange disorder. A large magnetic heat capacity $C_{\rm M}$  at low temperature is characteristic of these systems \cite{scgo,Silverstein2014,Clark2014,Plumb2017}, suggesting soft, gapless modes. The Goldstone modes described by the theory we develop here give rise to a large value for $C_{\rm M}$ because the excitation speed is small if exchange disorder is weak.

\section{Model and Ground States}
We study the classical Heisenberg model with Hamiltonian
\begin{align}
    {\cal H} &= \sum_{\braket{\bm{r},\bm{r'}}} J_{\bm{r},\bm{r'}} \bm{S}_{\bm{r}} \cdot \bm{S}_{\bm{r'}}\,.
\label{eq:H}
\end{align}
Here spins $\bm{S}_{\bm{r}}$ are three-component unit vectors, the sum is over nearest-neighbour pairs of sites $\bm{r},\bm{r'}$ on the pyrochlore lattice,  and  $J_{\bm{r},\bm{r'}} = J +  \Delta \cdot R_{\bm{r},\bm{r'}}$, where $R_{\bm{r},\bm{r'}}$ is a Gaussian random variable with zero mean and unit variance. Our focus is on the weak-disorder limit $\Delta \ll J$. 

In the absence of disorder ($\Delta=0$) this model has macroscopically degenerate ground states. These states are ones in which each tetrahedron $\alpha$ of the pyrochlore lattice has total spin ${\bm M}_\alpha \equiv \sum_{{\bm r}\in {\alpha}} {\bm S}_{\bm r}=0$, since ${\cal H} = (J/2)\sum_\alpha |{\bm M}_\alpha|^2+ {\rm constant}$. For $N$ spins under periodic boundary conditions, the number of ground-state degrees of freedom is $N/2+3$, a result which can be obtained simply (up to the finite-size term) by subtracting from the total number ($2N$) of degrees of freedom the number ($3N/2$) of scalar constraints implied by the conditions ${\bm M}_\alpha=0$\cite{Moessner1998a,Moessner1998}.

The ground states can be represented as configurations of an emergent gauge field as follows\cite{Isakov2004}. Noting that centres of tetrahedra lie on a bipartite lattice, one introduces unit real-space vectors with components $e^i_{\bm r}$ at each site $\bm r$, directed from one sublattice to the other. The emergent flux at site $\bm r$ is $B^{ai}_{\bm r} = S^a_{\bm r} e^i_{\bm r}$, where $a$ labels spin components and $i$ space components. The net flux into or out of a tetrahedron $\alpha$ is 
\begin{equation}\label{gs}
\sum_{{\bm r} \in {\alpha}} B^{ai}_{\bm r} e^i_{\bm r} = \sum_{{\bm r} \in {\alpha}} S_{\bm r}^a\, ,
\end{equation}
which is zero in ground states, so $B^{ai}_{\bm r}$ is divergenceless.

In the presence of weak disorder this model undergoes a spin freezing transition at low temperature \cite{Saunders2007,Andreanov2010}. The disorder realisation can be thought of as selecting a particular flux configuration from the ground state manifold of the clean system. By minimising the energy of a single cluster, and by direct numerical calculation, we find random ground state magnetisations of magnitude $|{\bm M}_{\alpha}| \sim \Delta/J$. These magnetisations represent canting away from the ground state manifold of the $\Delta = 0$ system, and vanish smoothly as $\Delta/J \to 0$ with fixed $R_{\bm{r,r'}}$. The frozen flux configurations, which are local minima of the energy landscape, are therefore smoothly connected to ground state flux configurations of the disorder-free model.

\section{Fluctuations}\label{fluctuations}
We are concerned for $\Delta \not= 0$ with small-amplitude excitations around a frozen spin configuration. As is standard in semiclassical treatments of fluctuations in Heisenberg magnets, these are characterised in two complementary ways: the energy costs of small-amplitude static fluctuations are represented by a Hessian; and the dynamics of fluctuations are given by normal modes of the linearised equations of motion. 

Let $\overline{\bm{S}}_{\bm{r}}$ denote a minimum energy configuration with energy $\overline{E}$. Write $\bm{S}_{\bm{r}}= (1-\frac{1}{2}\bm{m}_{\bm{r}}^2)\overline{\bm{S}}_{\bm{r}}+ \bm{m}_{\bm{r}}$ for $|\bm{m}_{\bm{r}}|\ll 1$ with $\bm{m}_{\bm r}\cdot \overline{\bm{S}}_{\bm{r}} =0$ at leading order. It is also useful to describe fluctuations in terms of a spin rotation $\bm{\theta}_{\bm r}$, by writing $\bm{m}_{\bm{r}} = \bm{\theta}_{\bm{r}} \times \overline{\bm{S}}_{\bm{r}}$. We take $\bm{\theta}_{\bm r} \cdot \overline{\bm{S}}_{\bm r} = 0$ so that $\bm{m}_{\bm{r}}$ and $\bm{\theta}_{\bm{r}}$ both have two dynamically conjugate degrees of freedom at each site. 

Static properties of excitations are characterised by the inverse susceptibility matrix $\chi^{-1}$ or Hessian. At quadratic order 
\begin{equation}
	{\cal H}-\overline{E} = \frac{1}{2} m^a_{\bm{r}} [\chi^{-1}]^{ab}_{\bm{r},\bm{r'}} m^b_{\bm{r'}} \equiv \frac{1}{2}\theta^a_{\bm{r}} \tau^{ab}_{\bm{r},\bm{r'}} \theta^b_{\bm{r'}}\,,
	\label{eq:chi}
\end{equation}
where the right-hand side of (\ref{eq:chi}) defines the matrix $\tau$. Let $\lambda_n$ for $n = 1 \ldots 2N$ denote the eigenvalues of the Hessian, and define their integrated density $N(\lambda) \equiv N^{-1}\sum_n \Theta(\lambda - \lambda_n)$, where $\Theta$ is the step function.

Dynamical properties follow from the precessional equation of motion
\begin{equation}
\partial_t \bm{S}_{\bm{r}} =  -\sum_{\bm{r'}} J_{\bm{r},\bm{r'}} \bm{S}_{\bm{r}} \times\bm{S}_{\bm{r'}}\,.
\end{equation}
After linearisation, this can be written as
\begin{align}
	\partial_t {m}^a_{\bm{r}} =  -\epsilon^{abc} \overline{{S}}^{b}_{\bm{r}} [\chi^{-1}]^{cd}_{\bm{r},\bm{r'}} {m}^d_{\bm{r'}}\,.
	\label{eq:linear_eom}
\end{align}
We denote the magnon eigenfrequencies by $\omega_n$ for $n=1 \ldots 2N$ and define their integrated density $D(\omega) \equiv N^{-1}\sum_n \Theta(\omega-|\omega_n|)$. Writing $m^a_{\bm{r}}(t) = m^a_{\bm{r}}e^{-i \omega t}$ in (\ref{eq:linear_eom}) and taking its complex conjugate, we see that eigenfrequencies come in $N$ pairs $\pm \omega$.

For $\Delta = 0$ the Hessian has $N/2+3$ zero eigenvalues, with the corresponding eigenvectors forming local coordinates for the ground-state manifold. The vanishing of $1/4$ (as $N\to\infty$) of the Hessian eigenvalues $\lambda$ reflects the extensive ground-state degeneracy. From the equation of motion, we see that the same fraction of dynamical eigenvalues are zero \cite{Moessner1998}. This has an important implication: it indicates that the coordinates within the ground state manifold, our emergent fluxes, form dynamically conjugate pairs. To see this, suppose that the opposite were true. Then some eigenvectors of the Hessian with zero eigenvalue would have canonically conjugate coordinates outside the ground-state manifold. Each such Hessian eigenvector would give rise to a pair of dynamical zero modes, and the fraction of dynamical zero modes would be larger than the fraction of Hessian zero modes.

Our approach to the system with weak exchange disorder is first to develop a picture of the Hessian eigenvectors, and then to use this to understand the dynamical modes. 
Using the result $|\bm{M}_{\alpha}| \sim \Delta/J$, we see that Eq.~(\ref{eq:chi}) takes the form
\begin{equation}
	\mathcal{H}-\overline{E} = (J/2)\sum_{\alpha}\Big(\sum_{\bm{r}\in \alpha}\bm{m}_{\bm{r}}\Big)^2 + \mathcal{O}(\Delta)\,.
\label{eq:chi2}
\end{equation}
In the regime $0 < \Delta \ll J$, consider fixing $R_{\bm{r,r'}}$ and letting $\Delta$ tend to zero. We can associate each local energy minimum of the spin-glass phase with an exact ground state of the clean system, in which $1/4$ of the $\lambda_n$ are zero. Increasing $\Delta$ from zero, and allowing the spin configuration to cant away from the ground state manifold, the only changes in expression (\ref{eq:chi2}) are of $\mathcal{O}(\Delta)$. Degenerate perturbation theory then shows that the soft fluctuations acquire stiffnesses $\lambda \sim \Delta$, the others being of order $J$. Therefore in the limit $\Delta \ll J$ we can separate the Hessian eigenvectors into subspaces with eigenvalues $\mathcal{O}(\Delta)$ and $\mathcal{O}(J)$, the first corresponding to fluctuations of the gauge fields, and the second to fluctuations of the cluster magnetisation.

This result for the energy cost of fluctuations of the emergent gauge fields, together with the fact that as $\Delta/J \to 0$ these degrees of freedom are dynamically conjugate, suggests that the lowest modes will be lifted from zero frequency to $\omega \sim \Delta$. If these are Goldstone modes, we see immediately that we are at odds with the conventional hydrodynamic theory of spin waves: a smooth magnetisation has an energy cost of order $J$ from (\ref{eq:chi2}), and the energy cost of a smooth rotation of the ground state configuration is at least of order $\Delta$. Consequently the conventional theory incorrectly predicts $\omega \sim \sqrt{\Delta J}$. To proceed, we develop a continuum treatment of the low-energy degrees of freedom of our model. These are related to smooth rotations of the frozen gauge field configuration. We then use this to construct a modified hydrodynamic theory of magnons, relevant to geometrically frustrated magnets with emergent gauge degrees of freedom and weak exchange disorder.

\section{Continuum Theory}\label{continuum}
We begin by examining how accurately a smooth spin rotation can be represented by the gauge-field degrees of freedom. Using a continuum treatment, let the tensor field $\overline{B}^{ai}({\bm r})$ denote a ground state selected by disorder in the limit $\Delta/J \to 0$. Note that Eq.~\eqref{gs} implies $\partial_i \overline{B}^{ai}({\bm r}) = 0$ and that divergenceful field configurations cost energy ${\cal O}(J)$. Rotations in spin-space can be described by an orthogonal matrix field $O^{ab}({\bm r})$ that satisfies $O^{ab}({\bm r}) O^{cb}({\bm r}) = \delta^{ac}$ for all $\bm r$. To ensure that a smoothly rotated configuration avoids an ${\cal O}(J)$ energy penalty, we write
\begin{equation}\label{smooth}
B^{ai}({\bm r}) = O^{ab}({\bm r}) \overline{B}^{bi}({\bm r})+ b^{ai}({\bm r})
\end{equation}
and choose $b^{ai}({\bm r})$ so that $\partial_i {B}^{ai}({\bm r}) = 0$. This implies
\begin{equation}\label{poisson}
\partial_i b^{ai}({\bm r}) = -[\partial_i O^{ab}({\bm r})] \overline{B}^{bi}({\bm r})\equiv \sigma^a({\bm r})\, ,
\end{equation}
and we show below that $b^{ai}({\bm r})$ is small for smooth rotations. Specifically, if $O^{ai}({\bm r})$ varies on a scale $\ell$, then we find that $|b^{ai}({\bm r})|^2 \sim \ell^{-3}$ for large $\ell$.

We would like an expression for the energy relative to the ground state of the configuration $B^{ai}({\bm r})$. Since uniform rotations cost no energy, it is natural to use a gradient expansion, and since we have imposed $\partial_i {B}^{ai}({\bm r}) = 0$, the associated stiffness is ${\cal O}(\Delta)$ and independent of $J$. Hence we write
\begin{equation}\label{continuumcost}
{\cal H} - \overline{E} \sim \Delta \int d^d \bm{r} \left\{ |\partial_i O^{ab}|^2 + \varepsilon(b)\right\}\, ,
\end{equation}
where $\varepsilon(b) \sim |b^{ai}(\bm{r})|^2$ characterises the energy density of corrections to the smooth rotation. Here we ignore disorder in the stiffness.

The energy density of the corrections scales as $\mathcal{O}(\ell^{-3})$. To see how this arises we solve \eqref{poisson} and write
\begin{align}
\int {\rm d}^3{\bm r} |b^{ai}({\bm r})|^2 =  \int \int{\rm d}^3{\bm r}_1 {\rm d}^3{\bm r}_2 \frac{\sigma^a({\bm r}_1)\sigma^a({\bm r}_2) }{4\pi |{\bm r}_1 - {\bm r}_2|}\,.
\end{align}
We now wish to substitute for $\sigma^a({\bm r}_1)$ in terms of $O^{ab}({\bm r})$ and $\overline{B}^{bi}({\bm r})$ using Eq.~\eqref{poisson}, and determine the dependence on the lengthscale $\ell$, which enters via the form chosen for $O^{ab}({\bm r})$. In this process the factor $\overline{B}^{bi}({\bm r}_1)\overline{B}^{ci}({\bm r}_2)$ appears. We replace it by its average \cite{Isakov2004}
\begin{equation}
\langle \overline{B}^{bi}(0)\overline{B}^{ci}({\bm r})\rangle \propto \delta^{bc} \frac{3r_i^2 - r^2}{r^5}\,
\label{eq:correlator}
\end{equation}
over a Gaussian ensemble of divergenceless fields, which can be justified by the averaging arising from integration over the centre of mass coordinate $\bm{r}_1+\bm{r}_2$. The $\ell^{-3}$ scaling then follows from power-counting \footnote{In two dimensions a similar approach gives $|b^{ai}({\bm r})|^2\sim \ell^{-2} \ln(\ell/a)$, and so in this case the constraint that $B^{ai}({\bm r})$ is divergenceless implies a stiffness that is logarithmically divergent at large $\ell$}. Therefore at large $\ell$ the leading contribution to the energy density associated with a smooth rotation of the gauge fields scales as $\Delta \ell^{-2}$ from Eq. \eqref{continuumcost}. We next show that these smooth rotations are intimately related to the dynamical Goldstone modes.

The low-frequency excitations involve an interplay between smooth rotations and the conserved magnetisation density. In a continuum description these are characterised by three-component vector fields $\bm{\theta}(\bm{r})$ and $\bm{m}(\bm{r})$, which are coarse-grained versions of their lattice counterparts, $\bm{\theta}_{\bm{r}}$ and $\bm{m}_{\bm{r}}$. Note however that although $\bm{\theta}_{\bm{r}}$ and $\bm{m}_{\bm{r}}$ provide equivalent descriptions of the spin fluctuations, this is no longer true after coarse-graining. For a conventional spin glass the continuum equations of motion proposed by Halperin and Saslow  \cite{HalperinB.I.;Saslow1977} are
\begin{equation}
	\bm{\dot \theta} = \chi_0^{-1} \bm{m} \quad {\rm and} \quad \bm{\dot m} =  \rho \nabla^2\bm{\theta},
\end{equation}
giving the value $c=\sqrt{\rho/\chi_0}$ for the speed.

To understand how this approach should be modified in the weakly disordered pyrochlore antiferromagnet, we start from the 
microscopic equation of motion (\ref{eq:linear_eom}), which can be recast in the two equivalent forms \cite{Ginzburg1978}
\begin{equation}
	\dot \theta^a_{\bm{r}} = [\chi^{-1}]^{ab}_{\bm{r}\bm{r'}} m^b_{\bm{r'}} \quad {\rm and} \quad \dot m^a_{\bm{r}} = -\tau^{ab}_{\bm{r}\bm{r'}} \theta^b_{\bm{r'}}.
\label{eq:double_eom}
\end{equation}
It is useful to expand a fluctuation $\bm{m}_{\bm r}$ in the basis of Hessian eigenvectors. As mentioned above, these span two subspaces, associated respectively with eigenvalues ${\cal O}(\Delta)$ and ${\cal O}(J)$. Separating the components of $\bm{m}_{\bm r}$ in each subspace, we write $\bm{m}_{\bm{r}}=\bm{m}_{\bm{r},\Delta}+\bm{m}_{\bm{r},J}$. Similarly, for $\bm{\theta}_{\bm r}$ in the basis of eigenvectors of $\tau$, we take $\bm{\theta}_{\bm{r}}=\bm{\theta}_{\bm{r},\Delta}+\bm{\theta}_{\bm{r},J}$. Since ground-state coordinates at $\Delta=0$ form dynamically conjugate pairs, each $\bm{\theta}_{{\bm r},\Delta}$ has a conjugate $\bm{m}_{{\bm r},\Delta}$.

Under coarse graining, only the smooth parts of $\bm{m}_{\bm r}$ and $\bm{\theta}_{\bm r}$ survive. These are contained in $\bm{m}_{\bm{r},J}$ and $\bm{\theta}_{\bm{r},\Delta}$. To see this, note first that the average $\sum_{\rm{r} \in \alpha} 
\bm{m}_{\bm{r},\Delta}$ over a tetrahedron $\alpha$ is zero for $\Delta/J \to 0$; hence $\bm{m}_{\bm{r},\Delta}$ is eliminated by coarse graining. Second, the expressions (\ref{eq:chi}) and (\ref{continuumcost}) for the energy, in terms of $\bm{\theta}_{\bm r}$ and $\bm{\theta}({\bm r})$ respectively, imply that smooth rotations are represented exclusively by $\bm{\theta}_{{\bm r},\Delta}$. The relevant coarse-grained degrees of freedom in a continuum theory are therefore $\bm{m}_J(\bm{r})$ and $\bm{\theta}_{\Delta}(\bm{r})$.

The coarse-grained equations of motion can be inferred from \eqref{eq:double_eom}. The first follows from the observation that a spin fluctuation $\bm{m}_{\bm{r},J}$ generates exchange fields of order $J$. This immediately yields
\begin{equation}\label{cont1}
	\bm{\dot \theta}_{\Delta}(\bm{r}) \sim J \bm{m}_J(\bm{r}).
\end{equation}
The second uses a comparison of the right-hand sides of \eqref{eq:chi} and \eqref{continuumcost} to establish that the action of $\tau$ on a smooth rotation 
$\bm{\theta}_{\bm{r},\Delta}$ 
can be represented in the continuum by $\tau \sim -\Delta a^2 \nabla^2$, where $a$ is the lattice spacing. The prefactor $\Delta$ sets the magnitude of the microscopic exchange fields generated by $\bm{\theta}_{\bm{r},\Delta}$, which in turn drive spin precession at frequency $\omega \sim \Delta$. As noted above, the canonically conjugate coordinate representing this precession can be written as $\bm{m}_{\bm{r},\Delta}$ in the limit $\Delta\to 0$. At finite $\Delta/J$ it is accompanied by a correction $\bm{m}_{\bm{r},J}$, with $|\bm{m}_{\bm{r},J}| \sim (\Delta/J) |\bm{m}_{\bm{r},\Delta}|$. Since these corrections alone survive coarse-graining, our second continuum equation of motion is
\begin{equation}\label{cont2}
	\bm{\dot m}_{J}(\bm{r}) \sim   a^2 \frac{\Delta^2}{J} \nabla^2 \bm{\theta}_{\Delta}(\bm{r}).
\end{equation}

In summary, a smooth magnetisation density, of magnitude ${\cal O}(\Delta/J)$ relative to the gauge field fluctuations, drives long-wavelength twists of the ground state configuration. From \eqref{cont1} and \eqref{cont2} we predict linearly dispersing Goldstone excitations with speed $c \sim a \Delta $ set only by exchange disorder. We next present numerical results that support this picture.

\begin{figure}
	\includegraphics[width=0.46\textwidth]{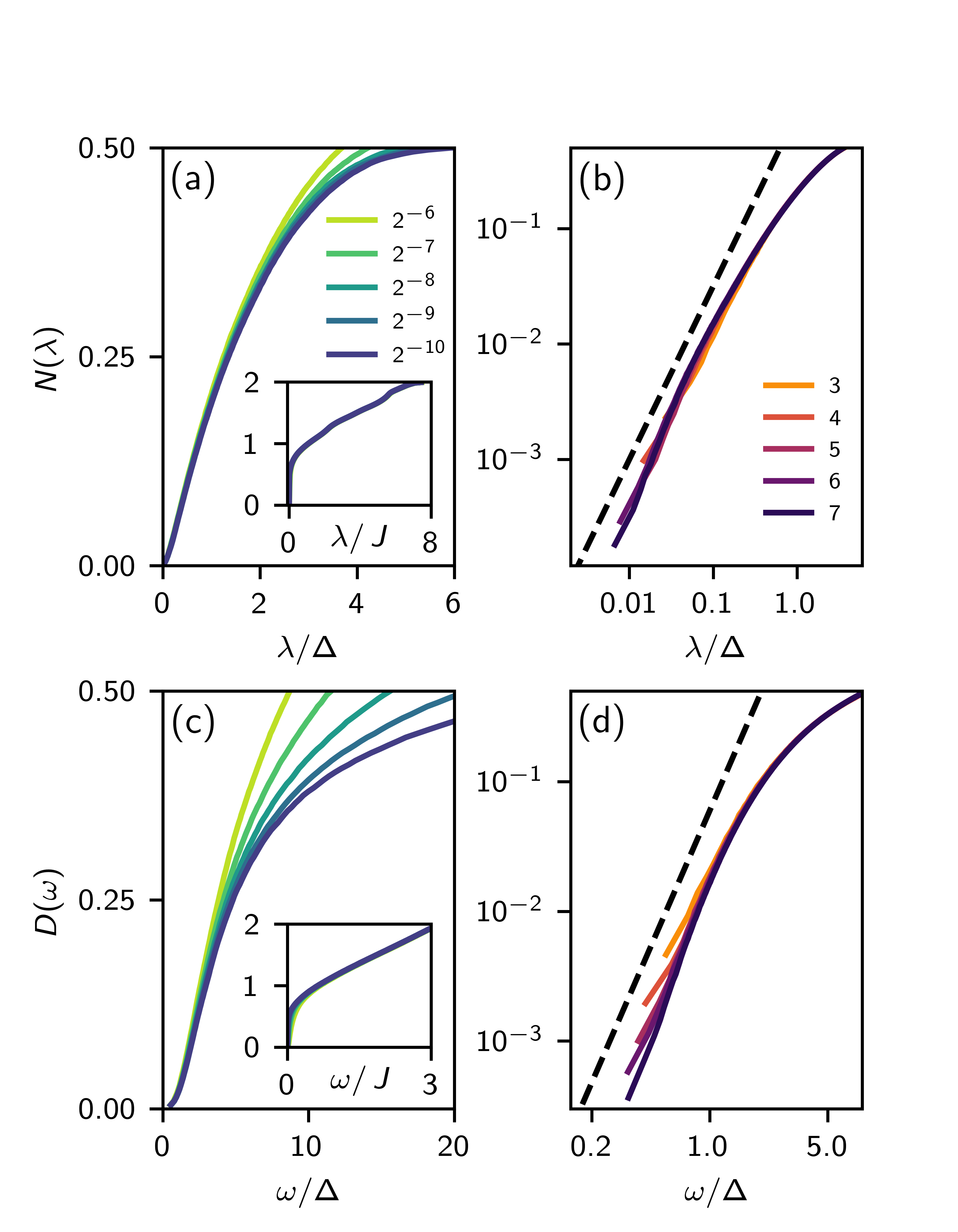}
	\caption{Integrated densities  $N(\lambda)$ of Hessian eigenvalues $\lambda$ [(a) and (b)] and $D(\omega)$ of magnon frequencies $\omega$ [(c) and (d)], with dependence on system size $L$ and disorder strength $\Delta/J$. Panels (a) and (c): for $2^{-10} \leq \Delta/J \leq 2^{-6}$ at $L=3$; panels (b) and (d): for $3\leq L \leq 7$ at $\Delta/J = 2^{-6}$. Dashed line in (b) represents $N(\lambda) \propto (\lambda/\Delta)^{3/2}$; dashed line in (d) represents $N(\omega) \propto (\omega/\Delta)^{3}$. Insets in (a) and (c) show full range of $N(\lambda)$ and $D(\omega)$.
	}
	\label{fig:combined}
\end{figure}

\begin{figure}
	\includegraphics[width=0.46\textwidth]{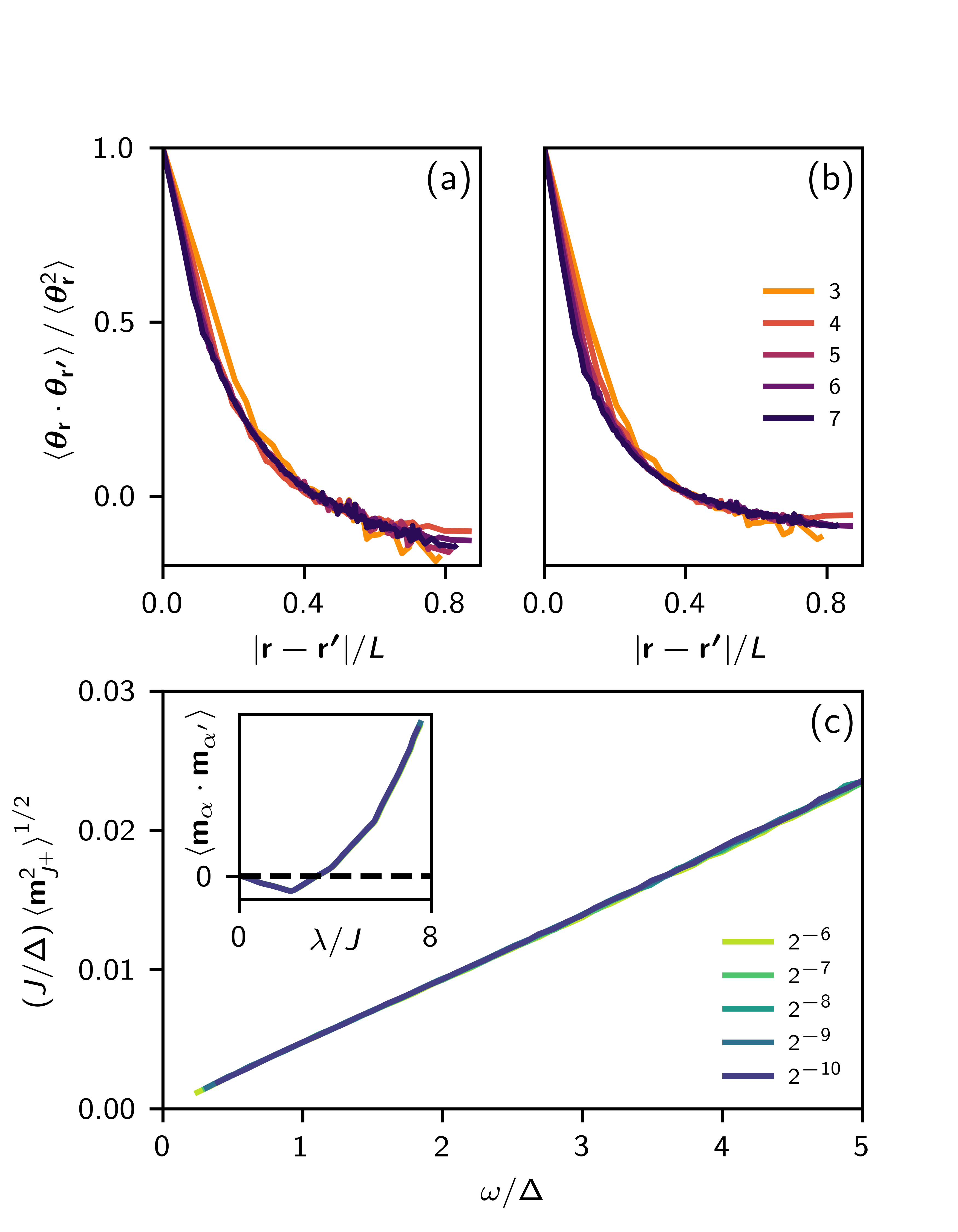}
	\caption{Characterisation of low-lying Hessian and dynamical eigenvectors. (a) Scaling collapse of angular correlator for lowest non-trivial Hessian eigenvectors, with $3\leq L \leq 7$ and $\Delta/J = 2^{-6}$. (b) Equivalent for dynamical eigenvectors. (c) Smooth magnetisation density in dynamical modes vs. $\omega/\Delta$ at $L=3$ for $2^{-10} \leq \Delta/J \leq 2^{-6}$. Inset: correlator of magnetisations of neighbouring tetrahedra in Hessian eigenvectors. See main text for discussion. Fluctuations evident in (a) and (b) for smaller $L$ are predominantly a finite-size effect, rather than because of sample-to-sample fluctuations.}
	\label{fig:theta}
\end{figure}

\section{Numerics}\label{numerics}
We generate low-lying minimum energy states  of $\cal H$ [Eq.~\eqref{eq:H}] by using a Metropolis algorithm to establish equilibrium at $T=0.1J$ and then reducing the energy via steepest descents \cite{WalkerL.R.;Walstedt1980}, iteratively rotating spins to be parallel to their local exchange fields $-\sum_{\bm{r'}}J_{\bm{r},\bm{r'}} \bm{S}_{\bm{r'}}$ with a maximum final error of $10^{-9}$ radians. We study cubic samples of linear dimension $L$ containing $N=16L^3$ spins. We average over $20$ disorder realisations for $L=3$ and over $10$ realisations for $4\leq L \leq 7$. In the following we discard the trivial eigenvectors of the Hessian and dynamical matrices related to global rotations \cite{WalkerL.R.;Walstedt1980}.

Results for Hessian eigenvalues $\lambda$ are presented in Fig.~\ref{fig:combined} (a) and (b), and those for the dynamical mode frequencies $\omega$ in Fig.~\ref{fig:combined} (c) and (d). The main panels of (a) and (c) show that, as $\Delta/J \to 0$, $1/4$ of eigenvalues or frequencies are ${\cal O}(\Delta)$; the insets to (a) and (c) show that the remainder are ${\cal O}(J)$. Fig.~\ref{fig:combined}(b) demonstrates that $N(\lambda) \propto (\lambda/\Delta)^{3/2}$ for $\lambda/\Delta$ and $\Delta/J$ small. This form follows from Eq.~\eqref{continuumcost} and a conventional mode-counting argument for translationally invariant systems with $\lambda$ quadratic in wavenumber.  Fig.~\ref{fig:combined}(d) shows $D(\omega) \propto (\omega/\Delta)^3$ for $\omega/\Delta$ and $\Delta/J$ small. This form follows from Eqns.~\eqref{cont1} and \eqref{cont2}, which imply linearly dispersing excitations.

Next we test our picture of low-lying Hessian eigenvectors and dynamical modes as long-wavelength twists of the ground-state spin configuration. In both cases we consider the lowest-lying mode that is not simply a global rotation, and start from the coordinates $\bm{\theta}_{\bm r}$. Since $\bm{\theta}_{\bm r}$ is defined to have no component along the equilibrium spin direction $\overline{\bm S}_{\bm r}$, it has spatial fluctuations even if it represents a global spin rotation. For this reason we redefine the coordinates to be $\bm{\theta}_{\bm{r}} = \overline{\bm{S}_{\bm{r}}} \times \bm{m}_{\bm r} + c_{\bm{r}} \overline{\bm{S}}_{\bm{r}}$, where $c_{\bm{r}}$ is determined by minimisation of $\sum_{\braket{\bm{r},\bm{r'}}}(\bm{\theta}_{\bm{r}}-\bm{\theta}_{\bm{r'}})^2$. This scheme ensures that in the case $\bm{m}_{\bm{r}} = \bm{\theta_0} \times \overline{\bm{S}}_{\bm{r}}$ we recover $\bm{\theta_r} = \bm{\theta_0}$ for all $\bm{r}$. In Fig.~\ref{fig:theta} (a) and (b) we present results for the connected correlator $\braket{\bm{\theta_{r}\cdot\theta_{r'}}}$ in the lowest non-trivial Hessian and dynamical modes, respectively. In both instances we find the scaling collapse $\braket{\bm{\theta_{r}\cdot\theta_{r'}}} = \braket{\bm{\theta}_{\bm r}^2}f(|\bm{r-r'}|/L)$ for data from system sizes $3 \leq L \leq 7$. This demonstrates that these modes predominantly involve twists of the minimum-energy spin configuration on the scale of the system size. 

Finally, we present evidence  in Fig.~\ref{fig:theta}(c) that the dynamics of $\bm{\theta}_{\bm{r},\Delta}$ for low-lying modes is, as argued in justification of \eqref{cont2}, driven by the smooth part of $\bm{m}_{\bm{r},J}$, which we denote by $\bm{m}_{J^+}$. Our expectation that $\omega \propto |\bm{m}_{J^+}|$ is vindicated by excellent scaling collapse  of $(J/\Delta) {\braket{\bm{m}_{J^+}^2}}^{1/2}$ vs. $\omega/\Delta$ for a range of $\omega/\Delta$ and $\Delta/J$. In this computation $\bm{m}_{J^+}$ is isolated by projecting $\bm{m}_{\bm{r},J}$ for a normalised dynamical eigenvector onto the subspace spanned by the Hessian eigenvectors associated with the highest quarter of $\lambda$. A simple indication that this subspace includes the smooth part of $\bm{m}_{\bm{r},J}$ is given in the inset to Fig.~\ref{fig:theta}(c), which shows the correlator $\braket{\bm{m}_{\alpha} \cdot \bm{m}_{\alpha'}}$ of magnetisations $\bm{m}_\alpha \equiv \sum_{\bm{r} \in \alpha} \bm{m}_{\bm r}$ of neighbouring tetrahedra $\alpha$ and $\alpha^\prime$ in Hessian eigenvectors. The correlator is positive in the subspace used to construct $\bm{m}_{J^+}$, as required if $\bm{m}_{\bm r}$ is smooth.

In conclusion, the data shown in Figs.~\ref{fig:combined} and \ref{fig:theta} provide extensive support for the theoretical results in sections \ref{fluctuations} and \ref{continuum}, and for the physical arguments used to derive them.  Most importantly, the data and physical arguments together establish the description of Goldstone modes in these systems as excitations of the emergent gauge fields.

\section{Discussion}
An experimental signature of these modes is their large magnetic contribution to the heat capacity, scaling as $(T/\Delta)^\alpha$ at $T \ll T_F$ with $\alpha=3$. Several examples of pyrochlore antiferromagnets show spin freezing at a temperature much lower than the main interaction scale, which is attributed to weak exchange disorder induced by random strains. Power-law magnetic heat capacity  (but with $\alpha \approx 2$) is reported for: NaCaNi$_2$F$_7$ (in which there is intrinsic disorder in the locations of non-magnetic cations) \cite{Plumb2017}; $\text{Y}_2 \text{Mo}_2 \text{O}_7$ \cite{Silverstein2014} (in which local lattice distortions have been detected \cite{Booth2000}); and  $\text{Lu}_2\text{Mo}_2\text{O}_2$ \cite{Clark2014}. It is also found in $\text{Sr}\text{Cr}_8\text{Ga}_4\text{O}_{19}$ \cite{scgo}; since this is a quasi-two dimensional material, the value $\alpha=2$ is expected here. We note that, since $D(\omega)$ [Fig.~\ref{fig:combined}(d)] is convex, there is an obvious reason for measured values of $\alpha$ to decrease as $T$ increases towards $T_F$. 

Inelastic neutron scattering would potentially provide more detailed information, although the small energy scales involved (smaller than $T_F$) present a challenge. Specifically, we expect that the energy-dependence of scattering from emergent gauge degrees of freedom should evolve with temperature, from a Lorentzian above $T_F$ to a triple-peaked form (elastic as well as gain and loss inelastic peaks)  below $T_F$. The wavevector dependence should be the same in all cases, with pinch-points and a suppression of scattering for small momentum transfer. 

The ideas we have developed are relevant to Heisenberg antiferromagnets that have emergent gauge fields as semiclassical, low-energy degrees of freedom. 
The gauge fields appear as zero-energy modes in a leading-order description of a system, such as the classical nearest-neighbour pyrochlore model we have considered. A Heisenberg spin-glass state arises at low temperature if the dominant correction to this leading-order description is from exchange randomness. Additional corrections, such as further neighbour interactions, do not impact our results provided they are sub-dominant; if they are dominant, a different (ordered) low-temperature state is expected.
Other frustrated magnets may require different treatments of weak disorder, an example being the jammed spin-liquid systems studied recently \cite{Bilitewski2017,Bilitewski2019}. Equally, in some contexts the dependence $c\sim \sqrt{\rho/\chi_0}$ may hold with distinct energy scales $\chi_0^{-1}$ and  $a^{-2}\rho$, as suggested \cite{Podolsky2009} for $\text{NiGa}_2\text{S}_4$ \cite{Nakatsuji2005}.

In summary, we have developed a theory of the Goldstone modes in a frustrated Heisenberg magnet with weak exchange randomness, illustrating how the standard hydrodynamic theory must be modified to understand their propagation. We find gapless excitations with energies depending only on the magnitude of the exchange randomness. 

We thank R. Moessner and A. Nahum for helpful discussions, and D. Podolsky for comments on the manuscript. This work was supported in part by EPSRC Grants EP/N01930X/1 and EP/S020527/1.

\bibliography{pyrobib}

\end{document}